\documentclass[aps,prd,preprintnumbers,twocolumn,amsmath,amssymb,superscriptaddress,showpacs,floatfix]{revtex4}
\usepackage{bbm}
\usepackage{dsfont}
\usepackage{mathrsfs}
\usepackage{amsmath,amssymb,bm,comment}
\usepackage{hyperref}
\usepackage{graphicx}
\usepackage{dcolumn}
\usepackage{bm}
\usepackage{color}

\begin{document}

\title{ Dirac sea and  chiral anomaly in the quantum kinetic theory}

\author{Jian-Hua Gao}
\affiliation{Shandong Provincial Key Laboratory of Optical Astronomy and Solar-Terrestrial Environment,
Institute of Space Sciences, Shandong University, Weihai, Shandong 264209, China}

\author{Zuo-Tang Liang}
\affiliation{Key Laboratory of Particle Physics and Particle Irradiation (MOE),
Institute of Frontier and Interdisciplinary Science,
Shandong University, Qingdao, Shandong 266237, China}

\author{Qun Wang}
\affiliation{Department of Modern Physics, University of Science and Technology
of China, Hefei, Anhui 230026, China}

\begin{abstract}
We revisit the chiral anomaly in the quantum kinetic theory in the Wigner function formalism
under the background field approximation.
Our results show that the chiral anomaly is actually from
the Dirac sea or the vacuum contribution in the un-normal-ordered Wigner function.
We also demonstrate that this contribution modifies the chiral kinetic equation for antiparticles.
\end{abstract}

\pacs{25.75.Nq, 12.38.Mh}

\maketitle

\section{Introduction}
The chiral anomaly is a novel and prominent quantum effect in particle physics and can only be understood in the quantum field level.
The kinetic theory is a bridge to connect the macroscopic physical magnitude with the microscopic particle motion in the classical phase space.
Hence it is a highly nontrivial task to incorporate the chiral anomaly into the kinetic approach in a consistent way
so that the chiral kinetic theory is properly formulated.
In recent years, a considerable amount of work on the chiral kinetic theory have been published
where the chiral kinetic equation has been derived from various methods such as
the semiclassical approach~\cite{Duval:2005vn,Wong:2011nt,Son:2012wh,Stephanov:2012ki,Dwivedi:2013dea,Chen:2014cla,Manuel:2014dza},
the Wigner function formalism~\cite{Gao:2012ix,Chen:2012ca,Hidaka:2016yjf,Huang:2018wdl,Gao:2018wmr,Liu:2018xip},
the effective field theory \cite{Son:2012zy,Carignano:2018gqt,Lin:2019ytz,Carignano:2019zsh}
and the world-line approach~\cite{Mueller:2017lzw,Mueller:2017arw,Mueller:2019gjj}.

In these publications, most works connect the chiral anomaly in the chiral kinetic theory with Berry's phase or Berry's curvature.
In contrast, it has also been pointed out by Fujikawa and collaborators~\cite{Deguchi:2005pc,Fujikawa:2005tv,Fujikawa:2005cn}
that topological effects due to Berry's phase and the chiral anomaly are basically different from each other.
The Berry phase arises only in the adiabatic limit while the chiral anomaly is generic  and independent of kinematic limits.
This distinct difference has been further demonstrated by Mueller and Venugopalan in~\cite{Mueller:2017lzw,Mueller:2017arw}
where they found out that the Berry's phase arises from the real part of the world-line effective action
in a particular adiabatic limit while the chiral anomaly is from the imaginary part.
In \cite{Hidaka:2017auj}, Hidaka and collaborators  performed the derivation of the chiral anomaly by using the non-trivial boundary condition of distribution function.
Recently, in our paper with other collaborators~\cite{Gao:2018wmr},
we found out that some singular boundary terms also result in a new source term
contributing to the chiral anomaly, in contrast to the well-known scenario of the Berry phase term.

It is clear that such a controversy situation needs to be further clarified
with a more fundamental approach.
In a very recent paper by Yee and Yi \cite{Yee:2019rot}, the authors
try to clarify the relationship between the chiral anomaly and the Chern number of the Berry connection via the conventional Feynman diagram.
In this paper, we try to clarify this situation from the quantum transport theory based on Wigner functions
~\cite{Heinz:1983nx,Elze:1986qd,Vasak:1987um,Zhuang:1995pd},
a first principle approach from quantum field theory.
To do this, we will start with the basic Wigner equations and derive the chiral anomaly step by step so that
one can see clearly where the contribution comes from.
The results obtained show that the Dirac sea or vacuum contribution originated from the anti-commutation relations
between antiparticle field operators in un-normal-ordered Wigner function can not be dropped casually.
This unique term that comes directly from the quantum field theory
plays a central role to generate  the right chiral anomaly in quantum kinetic theory both for massive and for massless fermion systems.
The coefficient of the chiral anomaly derived this way is universal and
is independent of the phase space distribution function at zero momentum.
We will also present the updated chiral kinetic equation for antiparticles with Dirac sea contribution.

The rest of the paper is organized as follows:
In Sec. II, we give a very brief review of the Wigner function formalism of the relativistic quantum kinetic theory
and present especially those equations that will be used in deriving the chiral anomaly.
In Sec. III, we derive the chiral anomaly from Wigner equations and show in particular that it is generated from the Dirac sea contribution.
In Sec. IV, we exhibit how the chiral kinetic equation is
derived from Wigner function formalism at chiral limit and give the updated chiral kinetic equation
for antiparticles with Dirac sea contribution. At last, we summarize the paper in Sec. VI.

\section{The Wigner function formalism}
\label{sec:Wigner}

We recall that the quantum kinetic theory based on Wigner functions is a first principle approach from quantum field theory.
Here, the Wigner matrix $W(x,p)$ is the basic unit, 
and for spin-1/2 fermions, 
is defined as the ensemble average of gauge invariant nonlocal bilinear Dirac spinor field,
\begin{equation}
\label{wigner}
W_{\alpha\beta} = \int\frac{d^4 y}{(2\pi)^4}
e^{-ip\cdot y}\langle \bar\psi_\beta(x_+)U(x_+,x_-) \psi_{\alpha}(x_-)\rangle,
\end{equation}
where   $x_{\pm}\equiv x \pm y/2$ are two space-time points centered at
$x$ with separation $y$, and $U$ denotes the gauge link along the straight line  between $x_+$ and $x_-$,
\begin{equation}
\label{link}
U (x_+,x_- ) \equiv e^{-i  \int_{x_-}^{x_+} dz^\mu A_\mu (z)}.
\end{equation}
Here we did not define the  ensemble average  with normal ordering for the Dirac fields  because the Wigner equation (\ref{eq-c}) below  derived  from Dirac equation
 must be satisfied by the Wigner function without normal ordering instead of with normal ordering. Besides, we  did not include the path-ordering in the definition of gauge link above
since we  restrict ourselves to  background field approximation in this work.
The electric charge has been absorbed into the gauge potential $A_\mu$ for brevity.
Under the background field approximation, we obtain the equation satisfied by the Wigner matrix
from Dirac equation as given by~\cite{Vasak:1987um},
\begin{equation}
\label{eq-c}
\gamma_\mu ( \Pi^\mu +\frac{i}{2} G^\mu -m)  W(x,p)=0 ,
\end{equation}
where $\gamma^{\mu}$'s are Dirac matrices and $m$ is the particle's mass. The operators $G^\mu$ and $\Pi^\mu$ denote
the non-local generalizations of the space-time derivative $\partial^\mu_x$ and the momentum $p^\mu$,
\begin{eqnarray}
\label{nabla-mu-c}
G^\mu &\equiv& \partial^\mu_x- j_0(\frac{1}{2}\hbar\Delta)F^{\mu\nu}\partial_\nu^p,\\
\label{Pi-mu-c}
\Pi^\mu &\equiv& p^\mu -\frac{1}{2}\hbar j_1(\frac{1}{2}\hbar\Delta)F^{\mu\nu}\partial_\nu^p,
\end{eqnarray}
where $j_0$ and  $j_1$ are the zeroth and the first order spherical Bessel functions, respectively.
The triangle operator is defined as $\Delta\equiv \partial^p\cdot \partial_x$,
in which $\partial_x$  acts only on the field strength tensor $F^{\mu\nu}$ but not on the Wigner function.

We emphasize once more that in the definition of the Wigner function given by Eq.~(\ref{wigner})
and the Wigner equation given by Eq.~(\ref{eq-c})
there is no normal ordering in the Wigner matrix.
We will illustrate that this point plays a central role to give rise to the chiral anomaly in the quantum kinetic theory.
The Hamiltonian derivation of the chiral anomaly from the physically correct normal ordering can be found in~\cite{Dunne:1989gp}.

The Wigner equation given by Eq.~(\ref{eq-c}) is a matrix equation
that actually include 32 equations for 16 independent components in Wigner matrix.
These 16 components are classified as
scalar $\mathscr{F}$, pseudo-scalar $\mathscr{P}$, vector $\mathscr{V}_\mu$,
axial-vector $\mathscr{A}_\mu$ and tensor $\mathscr{S}_{\mu\nu}$ components according to the Lorentz transformation.
They are all real functions of $x$ and $p$ defined by the $\Gamma$-matrix expansion of $W(x,p)$, i.e.,
\begin{eqnarray}
\label{decomposition}
W=\frac{1}{4}(\mathscr{F}+i\gamma^5 \mathscr{P}+\gamma^\mu \mathscr{V}_\mu +\gamma^5 \gamma^\mu \mathscr{A}_\mu
+\frac{1}{2}\sigma^{\mu\nu} \mathscr{S}_{\mu\nu}).
\end{eqnarray}

The chiral anomaly is a quantum effect that can be studied using the semiclassical expansion in terms of $\hbar^n$.
It has been shown that~\cite{Gao:2019znl},  for massive particles,
up to the first order of $\hbar$ in the expansion,  
we can choose $\mathscr{F}$ and $\mathscr{A}_{\mu}$  as independent dynamical Wigner functions
and sort the 32 Wigner equations as follows:
Eleven of them provide explicit expressions of other Wigner functions in terms of $\mathscr{F}$ and $\mathscr{A}_{\mu}$,
\begin{eqnarray}
\label{P-2-b}
\mathscr{P}&=&-\frac{\hbar}{2m} \nabla^\mu \mathscr{A}_{\mu}, \\
\label{V-1-b}
\mathscr{V}_{\mu} &=& \frac{p_\mu}{m}  \mathscr{F}
-\frac{\hbar}{2m^2}\epsilon_{\mu\nu\rho\sigma}\nabla^\nu p^\rho \mathscr{A}^{\sigma}, \\
\label{S-1-b}
\mathscr{S}_{\mu\nu} &=&-\frac{1}{m} \epsilon_{\mu\nu\rho\sigma}p^\rho \mathscr{A}^{\sigma}
+\frac{\hbar}{2m^2}\left( \nabla_{\mu} p_{\nu} - \nabla_{\nu} p_{\mu}\right) \mathscr{F},~~~
\end{eqnarray}
five of them give transport equations, 
\begin{eqnarray}
\label{Veq-1-b}
p^\mu \nabla_\mu\mathscr{F}&=&\frac{1}{2m}p^\mu \Delta \tilde F_{\mu\nu} \mathscr{A}^{\nu},\\
\label{PS-2-b}
p^\nu \nabla_\nu  \mathscr{A}_{\mu} &=&F_{\mu\nu} \mathscr{A}^{\nu}
+\frac{1}{2m}p^\nu \Delta \tilde F_{\mu\nu}\mathscr{F},
\end{eqnarray}
where $\nabla^\mu \equiv \partial^\mu_x- F^{\mu\nu}\partial_\nu^p$, $\epsilon^{0123}=1$,
and $\tilde F_{\mu\nu}=\epsilon_{\mu\nu\alpha\beta}F^{\alpha\beta}/2$;
and another five provide mass shell equations 
with the following general solutions,
\begin{eqnarray}
\label{F-1-b-3v}
\mathscr{F}&=&\delta(p^2-m^2)\mathcal{F}+\frac{\hbar}{m} \tilde F_{\mu\nu} p^\mu \mathcal{A}^{\nu}\delta^{\prime}(p^2-m^2),\\
\label{A-1-b-3v}
\mathscr{A}_{\mu} &=& \delta(p^2-m^2)  \mathcal{A}_{\mu} +\frac{\hbar}{m} \tilde F_{\mu\nu} p^\nu\mathcal{F}\delta^{\prime}(p^2-m^2),
\end{eqnarray}
where  $\mathcal{F}$ and $ \mathcal{A}_{\mu}$ are arbitrary functions that are non-singular for on-shell momentum $p^2=m^2$.
They can be taken as the fundamental functions replacing $\mathscr{F}$ and $\mathscr{A}_{\mu}$ in practice.
Finally, there is another constraint equation for the axial vector component $\mathscr{A}_{\mu}$,
\begin{eqnarray}
\label{Acon-1-b}
 p^\mu\mathscr{A}_\mu &=&0.
\end{eqnarray}
and all other 10 Wigner equations are satisfied automatically thus are not needed to consider in practice.

We will now use these results to derive the chiral anomaly in the following. Other work on
quantum kinetic theory for massive fermion can be found in Refs.\cite{Chen:2013iga,Fang:2016uds,Weickgenannt:2019dks,Hattori:2019ahi,Wang:2019moi,Li:2019qkf}.

\section{The Chiral anomaly}
\label{sec:chiral}

Using the results presented in Sec.~\ref{sec:Wigner}, we can calculate the chiral anomaly to the first order in $\hbar$.
In the following of this section, we present the calculations step by step.
To do this, we start with the expression of $\mathscr{P}$ given by Eq. (\ref{P-2-b}).
By inserting the general solution of the mass shell equation for $\mathscr{A}_{\mu}$ given by Eq.~(\ref{A-1-b-3v}) into Eq.(\ref{P-2-b})
and integrating over the four-momentum $p$, we obtain the divergence of the axial current $j_\mu^{5}$ as,
\begin{eqnarray}
\label{chiral_anomaly}
\hbar \partial^\mu_x j_\mu^{5}= - 2m j_5 + \hbar X - \frac{\hbar^2}{8\pi^2} C F^{\mu\nu} \tilde F_{\mu\nu},
\end{eqnarray}
where $j_\mu^{5}=\int d^4 p \mathscr{A}_{\mu}$, $j_5 = \int d^4 p \mathscr{P}$ and
\begin{eqnarray}
X &=& F^{\mu\lambda}\int d^4 p \partial_\lambda^p\left[  \mathcal{A}_{\mu}\delta\left(p^2-m^2\right) \right],\\
\label{CA_C}
C &=& - \frac{2\pi^2}{m}\int d^4 p \, \partial^\lambda_p \left[ p_\lambda  \mathcal{F}\delta^{\prime}\left(p^2-m^2\right)\right].
\end{eqnarray}
Though the presence of $m$ in the denominator in the expression of $C$ given by Eq.~(\ref{CA_C}),
Eq.~(\ref{chiral_anomaly}) in the chiral limit $m=0$ is finite because $\mathcal{F} \propto m$
and the result turns out to be identical to  the one derived directly from
the chiral kinetic theory~\cite{Gao:2012ix,Chen:2012ca,Gao:2018wmr}.
Therefore, Eq.~(\ref{chiral_anomaly}) actually holds for both massive and massless fermions.

We now calculate these two coefficients $X$ and $C$ carefully to find out the source of the chiral anomaly
by using the solution of $\mathcal{A}_{\mu}$ and $\mathcal{F}$.
First of all, it is easy to verify that $X$ vanishes if $\mathcal{A}_{\mu}$ approaches zero rapidly at infinity in momentum space.
Such a result is obvious since the chiral anomaly is quantum effect and should not exist in the classical limit.
This boundary condition seems reasonable for $\mathcal{A}_{\mu}$ from the result at zeroth order that can be calculated directly
from the free quantum field theory, i.e.,
\begin{eqnarray}
\label{eq:Afree}
{\mathcal{A}}_\mu &=& \left\{
                       \begin{array}{ll}
                              \frac{ m}{4\pi^3} s_\mu (f^+ - f^-),\ \ \ p_0>0 \\
                              \     \\
                              \frac{ m}{4\pi^3}s_\mu (\bar f^+ - \bar f^-),\ \ \  p_0<0.
                       \end{array}
                     \right.
\end{eqnarray}
 where $s_\mu$ is the spin polarization vector with $s^2=-1$ and $s\cdot p=0$,
 and $f^\pm$ and $\bar f^\pm$ are number densities in the phase space for particles
and antiparticles with spins $\pm s_\mu$, respectively.
They are defined as the ensemble averages of the normal-ordered number density operators and
are expected to vanish at infinity in the phase space.  For ${\mathcal{A}}_\mu$, there is no non-trivial Dirac sea or
vacuum contribution from  the anticommutator of the Dirac field.

The situation is however different for the coefficient $C$ where solution of the scalar function ${\mathcal{F}}$ is needed.
Here at the same level as $\mathcal{A}_{\mu}$ given by Eq.~(\ref{eq:Afree}),
we have the result from the free quantum field theory as,
%
\begin{eqnarray}
\label{Fpm}
{\mathcal{F}}&=& \left\{
                       \begin{array}{ll}
                              \frac{m}{4\pi^3}( f^+ + f^-),\ \ \ p_0>0 \\
                              \     \\
                              \frac{ m}{4\pi^3}(\bar f^+ + \bar f^- + 2\bar f_{\textrm{v}}),\ \ \  p_0<0.
                       \end{array}
                     \right.
\end{eqnarray}
We see that, in contrast to the axial vector ${\mathcal{A}}_\mu$,
there exists a Dirac sea or vacuum contribution $\bar f_{\textrm{v}}=-1$ to the scalar function ${\mathcal{F}}$ \cite{Sheng:2017lfu,Sheng:2018jwf}.
This contribution originates from the anticommutator of the antiparticle field in the definition
of Wigner function given by Eq.~(\ref{wigner}) without normal ordering.
It takes the value $\bar f_{\textrm{v}}=-1$ that is universal and does not depend on the state of the system that we are considering.

Now we show how the universal $\bar f_{\textrm{v}}=-1$ gives rise to the universal coefficient $C$ of the chiral anomaly in front of
$F^{\mu\nu} \tilde F_{\mu\nu}$ in Eq.~(\ref{chiral_anomaly}).
We consider in Eq.~(\ref{Fpm}) only this universal Dirac sea contribution $\bar f_{\textrm{v}}=-1$ and substitute it into Eq.~(\ref{CA_C}) and obtain,
\begin{eqnarray}
\label{P-2-v-int0}
C_{\textrm v}
&=& \int\frac{ d^4 p}{\pi} \partial_{p_0} \left[
\frac{1}{4E_p}\delta'(p_0+E_p)\right]\nonumber\\
& &- \int \frac{d^4 p}{\pi}{\mathbf \partial}_{\bf p}\cdot \left[ {\bf p}\,
\frac{1}{4E_p^2}\delta'(p_0+E_p)\right]\nonumber\\
& &+ \int \frac{d^4 p}{\pi}{\mathbf \partial}_{\bf p}\cdot \left[ {\bf p}\,
\frac{1}{4E_p^3}\delta(p_0+E_p)\right],
\end{eqnarray}
where $E_p=\sqrt{{\bf p}^2+m^2}$ and we use the subscript ``v" to denote the Dirac sea contribution only.
It is easy to verify that the first and second terms vanish by direct calculation while the last term survives and gives,
\begin{eqnarray}
\label{CA-BC}
C_{\textrm v}&=& \int \frac{d^3{\bf p}}{2\pi}{\mathbf \partial}_{\bf p}\cdot \left( \,
\frac{{\bf p}}{2E_p^3}\right)=1.
\end{eqnarray}
This is just the right coefficient of the chiral anomaly.
It is also obvious that this derivation does not depend on whether the system is for massive or massless fermions.
At the chiral limit $m=0$ and $E_p=|{\bf p}|$,
hence ${{\bf p}}/{ (2 E_p^3)}$ in the bracket in Eq.~(\ref{CA-BC}) is just the usual Berry curvature
${\pmb \Omega}={\bf p}/(2 |\bf p|^3)$ with ${\pmb \nabla}\cdot {\pmb \Omega}=2\pi$,
and this exhibits how the chiral anomaly comes from the Berry curvature.
As we all know, we can carry out the integration above either directly in
the 3-dimensional momentum volume where the Berry monopole appears and only the infrared momentum contributes,
or in the 2-dimensional momentum boundary area by using the Gauss theorem in which only the ultraviolet momentum contributes.
This illustrates how the ultraviolet and infrared regions in momentum space are connected.
However, for the massive particle there is no infrared singularity or Berry monopole.
When the Dirac sea contribution is included, the boundary conditions of  distribution functions
at infinity of momentum space for particle and antiparticle are different and
this will be compatible with the treatment given in~\cite{Hidaka:2017auj}.

Now we analyze the normal contributions associated with $f^{\pm}$ and $\bar f^{\pm}$.
Totally, they give no contribution to the chiral anomaly
because usually the normal distribution functions $f^{\pm}$ and $\bar f^{\pm}$
are all supposed to vanish rapidly  at infinity in the phase space
so that they lead to no contribution after the integration in Eq.~(\ref{CA_C}).
The details of calculations are given in the Appendix.

We note that although we relate the chiral anomaly with the Berry curvature through Eq.~(\ref{CA-BC}) at chiral limit,
it is different from the result given in Refs.~\cite{Stephanov:2012ki,Manuel:2014dza,Chen:2012ca}
where the coefficient of the chiral anomaly is generated from
\begin{eqnarray}
\label{C-berry-1}
C&=&-\int\frac{ d^3{\bf p}}{2\pi}
 {\pmb \Omega} \cdot {\mathbf \partial}_{\bf p} (f_{\bf p}+\bar f_{\bf p})
\end{eqnarray}
After integrating by parts and using ${\pmb \nabla}\cdot {\pmb \Omega}=2\pi$, we obtain
\begin{eqnarray}
\label{C-berry-2}
C&=& f_{{\bf p}=0}+\bar f_{{\bf p}=0}
\end{eqnarray}
Hence it depend on the specific distribution function at infrared momentum. It is interesting that with the Fermi-Dirac distribution,
\begin{eqnarray}
\label{Fermi-Dirac}
f^{\pm}&=&\frac{1}{e^{( |{\bf p}|-\mu_{\pm})/T}+1},\\
\bar f^{\pm} &=& \frac{1}{e^{( |{\bf p}| + \mu_{\pm})/T}+1}
\end{eqnarray}
the Eq.(\ref{C-berry-2}) gives the right coefficient of the chiral anomaly.
In our approach, the similar term to Eq.(\ref{C-berry-1})  indeed exist as well and is included in
$C_\textrm{n}$  where the subscript ``n'' indicates that only the normal distribution function $f^{\pm}$ and $\bar f^{\pm}$ are involved, see the definition in Eq.~(\ref{P-2-n-int0}) in Appendix.
However they always appear in the divergence form, see the first term $C_\textrm{n}^{(1)}$ of $C_\textrm{n}$ given by Eq.~(\ref{Cn1}),
\begin{eqnarray}
\label{Cn1-final-a}
C_\textrm{n}^{(1)}&=&-\int\frac{ d^3{\bf p}}{4\pi} {\mathbf \partial}_{\bf p}\cdot
\left[ \frac{{\bf p} }{ E_p^3} (f_{\bf p}+\bar f_{\bf p})\right] \nonumber\\
&=&-\int\frac{ d^3{\bf p}}{4\pi} \frac{{\bf p}  }{ E_p^3} \cdot {\mathbf \partial}_{\bf p}(f_{\bf p}+\bar f_{\bf p})\nonumber\\
& &-\int\frac{ d^3{\bf p}}{4\pi}(f_{\bf p}+\bar f_{\bf p}) {\mathbf \partial}_{\bf p}\cdot
\left( \frac{{\bf p} }{ E_p^3}\right),
\end{eqnarray}
where $f_{\bf p}$ and $\bar f_{\bf p}$ in our approach  are defined by Eq.~(\ref{def:f}).
We see that $C_\textrm{n}^{(1)}$ is divided into two terms and
the sum of them vanishes for $f_{\bf p}$ and $\bar f_{\bf p}$ that go to zero at infinity of momentum.
At chiral limit $m=0$ and with specific Fermi-Dirac distribution,   the first term  in $C_\textrm{n}^{(1)}$ reproduces the result (\ref{C-berry-2}) and give the coefficient of the chiral anomaly.
However the second term always cancel this term and  the total contribution must vanish.
Hence in such very special case, one might regard the Berry's curvature
as the effective source for the chiral anomaly because the Dirac sea contribution
and the second term in Eq.~(\ref{Cn1-final-a}) happens to cancel with each other.
However, 
such a case is a coincidence since it holds only for the chiral system
with a very specific distribution constraint $f_{\bf p}+\bar f_{\bf p}=1$ at the zero momentum point. It does not hold for massive fermions at all.
In general, the Berry's curvature contribution to the chiral anomaly is always cancelled by
the second term in Eq.~(\ref{Cn1-final-a}) and
the real contribution to the chiral anomaly comes actually from Dirac sea $C_\textrm{v}$.

Now let us consider another interesting term --- the last term $C_\textrm{n}^{(3)}$ in Eq.~(\ref{P-2-n-int0}).
At the chiral limit and for the isotropic distribution function in the momentum space, we have,
\begin{eqnarray}
\label{Cn1-final}
C_{\textrm n}^{(3)}&=&
\left( f_{p_0=0}+\bar f_{p_0=0}\right) -\left( f_{p_0=0}+\bar f_{p_0=0}\right).
\end{eqnarray}
Once more, if we choose the Fermi-Dirac distribution,
the first term gives the right coefficient ``1'' of the chiral anomaly though it is always cancelled by the second term.

To summarize, we see that for massless Fermion with special distribution functions
$f_{\bf p}+\bar f_{\bf p}=1$ and $f_{p_0=0}+\bar f_{p_0=0}=1$,
we can rewrite Eq.~(\ref{chiral_anomaly}) as,
\begin{eqnarray}
\partial^\mu_x j_\mu^{5(1)} &=&- \frac{1}{8\pi^2}
\left[ C_{\textrm n}^{(1)}+C_{\textrm n}^{(3)}+{C_{\textrm v}}\right] F^{\mu\nu}\tilde F_{\mu\nu}\nonumber\\
&=&- \frac{1}{8\pi^2} \left[1-1+1-1+{1}\right] F^{\mu\nu}\tilde F_{\mu\nu}\nonumber\\
&=& -\frac{1}{8\pi^2} F^{\mu\nu}\tilde F_{\mu\nu}.
\end{eqnarray}
In our previous work in \cite{Gao:2018wmr}, the first, third and fourth terms were obtained while the second term and last term were missing.
Some other work kept  only third term corresponding to Berry curvature term.

\section{The Chiral Kinetic Equation}
\label{sec:CKE}
In this section, we show that the Dirac sea contribution modifies the chiral kinetic equation for the antiparticle.
We now first recall how we obtain the chiral kinetic equation from the Wigner function approach.
In the chiral limit, it is convenient to define the helicity basis,
\begin{equation}
\mathscr{J}^{\mu}_{s}=\frac{1}{2}\left(\mathscr{V}^{\mu}+s\mathscr{A}^{\mu}\right),
\label{vwfc}
\end{equation}
where $s=\pm$ is the chirality.
In this chiral limit, the helicity basis are completely decoupled from each other and
from all the other Wigner functions as well,
\begin{eqnarray}
p^{\mu}\mathscr{J}_\mu & = & 0,\nonumber \\
\nabla^{\mu}\mathscr{J}_\mu & = & 0,\nonumber \\
2s\left( p_{\mu}\mathscr{J}_{\nu} -p_{\nu}\mathscr{J}_{\mu}\right) & = & -\hbar\epsilon_{\mu\nu\rho\sigma}\nabla^{\rho}\mathscr{J}^{\sigma},
\label{eq:wig-eq-1}
\end{eqnarray}
where we have suppressed the lower index ``$s$'' for brevity and only kept the contribution up to the first order of $\hbar$.

As have shown in~\cite{Gao:2018wmr}, in the chiral limit, a disentanglement theorem of Wigner function is valid.
According this theorem, only one of the four components of the Wigner function $\mathscr{J}_\mu$ is independent,
all the other three can be determined completely from it.
This independent Wigner function satisfies only one Wigner equation
and the other equations are satisfied automatically.
Especially we have the freedom to choose which component as the independent one.
In general, we can introduce a time-like 4-vector $n^\mu$ with normalization $n^2=1$
and choose $\mathscr{J}_n$ as the independent Wigner function.
For simplicity we assume $n^\mu$ is a constant vector.
With the auxiliary vector $n^\mu$, we can  decompose any vector $X^\mu$ into
the component parallel to $n^\mu$ and that perpendicular to $n^\mu$,
\begin{equation}
X^\mu=X_n n^\mu + \bar X^\mu,
\label{decomp-n}
\end{equation}
where $X_n =n\cdot X$ and $\bar X \cdot n=0$.
The electromagnetic tensor $F^{\mu\nu}$ can be decomposed  into,
\begin{equation}
F^{\mu\nu}=E^\mu n^\nu -E^\nu n^\mu +\epsilon^{\mu\nu\rho\sigma}n_\rho B_\sigma.
\end{equation}
With $ \mathscr{J}_n$ as the independent Wigner function, we obtain the other components,
\begin{eqnarray}
 \bar{ \mathscr{J}}_{\mu} &=& \bar p_\mu\frac{ {\mathscr{J}}_{n} }{ p_n }
-\frac{s\hbar }{2 p_n} \epsilon^{\mu\nu\rho\sigma} n_\nu  \nabla _{\sigma} \left(p_\rho \frac{ {\mathscr{J}}_{n} }{ p_n }\right),
\label{j0-j1-cov}
\end{eqnarray}
by contracting both sides of the last equation in Eq.~(\ref{eq:wig-eq-1}) with $n^\nu$.
Substituting this relation into the first equation of Eq.~(\ref{eq:wig-eq-1}),
we obtain the general expression,
\begin{eqnarray}
\frac{{\mathscr{J}}_{n}}{  p_n } &=& f \delta\left(p^2\right)
- \frac{s\hbar  }{p_n} B\cdot p f \delta^{\prime}\left(p^2\right) .
\end{eqnarray}
It follows that,
\begin{eqnarray}
{\mathscr{J}}_{\mu}
&\approx&\left( g_{\mu\nu} + \frac{\hbar s}{2p_n} \epsilon_{\mu\nu \rho \sigma}  n^\rho   \nabla^\sigma  \right)\nonumber\\
& &\times\left[ p^\nu f \delta\left(p^2-\hbar s \frac{B\cdot p}{p_n}\right)\right].\hspace{0.7cm}
\end{eqnarray}
Inserting this result into the second equation in Eq.~(\ref{eq:wig-eq-1}), we obtain the covariant chiral kinetic equation
up to the first order of $\hbar$ as,
\begin{eqnarray}
& &\nabla^\mu \left\{\left( g_{\mu\nu} + \frac{\hbar s}{2p_n} \epsilon_{\mu\nu \rho \sigma}  n^\rho   \nabla^\sigma  \right)\right.\nonumber\\
& &\left.\times\left[ p^\nu f \delta \left(p^2-\hbar s \frac{B\cdot p}{p_n}\right)\right]\right\}=0.
\end{eqnarray}
Integrating over $p_n$ from $0$ to $+\infty$ gives rise to the chiral kinetic equation for particles,
\begin{eqnarray}
\label{cke-particle}
 \left(1 + s\hbar B\cdot \Omega\right)  n\cdot \partial^x  f_{\bar p}& &\nonumber\\
+ \left[v^\mu + s\hbar ( \hat{\bar p} \cdot \Omega ) B^\mu
+ {s\hbar}\epsilon^{\mu\nu \rho \sigma} n _\rho E_\sigma \Omega_\nu\right] \bar \partial^x_\mu f_{\bar p} & &\nonumber\\
+\left( \tilde E^{\mu} + \epsilon^{\mu\nu\alpha\beta} v_{\nu}  n_\alpha B_\beta
 + {s\hbar} E\cdot B    \Omega^\mu \right)\bar \partial_\mu^p   f_{\bar p} & &  \nonumber\\
 + {s\hbar}E\cdot B \left( \bar\partial^\mu_p \Omega_\mu\right) f_{\bar p} &=&0, \hspace{1cm}
\end{eqnarray}
where
\begin{eqnarray}
 f_{\bar p}&=&f\left({p_n=|\bar p|(1+\hbar s B\cdot \Omega)}\right), \\
\Omega^\mu &=& \frac{  \bar p^\mu   }{ 2|\bar p|^3},\ \ |\bar p|= \sqrt{-\bar p^2},\ \ \hat {\bar p}_\mu=\frac{\bar p_\mu}{|\bar p|},\\
v^\mu &=&  \left(1 + \frac{s\hbar B\cdot \bar p}{|\bar p|^3}\right)\hat{\bar p}^\mu + \frac{s\hbar B^\mu}{2|\bar p|^2},\\
\tilde E^\mu &=& E^{\mu} -{s\hbar} (\bar \partial^\mu_x B^\lambda) \Omega_\lambda.
\end{eqnarray}
For finite momentum $\bar p^\mu$, the last term in Eq.~(\ref{cke-particle}) vanishes due to the Berry's monopole in momentum space
 $\partial_\mu^p \Omega^\mu = 2\pi \delta^3(\bar p)$
and the chiral kinetic equation reduces to the usual form obtained in  \cite{Stephanov:2012ki,Son:2012zy,Manuel:2014dza,Chen:2012ca,Hidaka:2016yjf,Huang:2018wdl,Gao:2018wmr} and so on.
However, once we leave the infrared region where the last term can be neglected,
the place where the chiral anomaly comes from is also concealed,
because this last term will always cancel the last term in the second line from below in Eq.~(\ref{cke-particle})
after integrating over the momentum.

By integrating over $p_n$ from  $-\infty$ to $0$ and replacing  $\bar p$  and $s$ with  $ -\bar p$ and $-s$ respectively,
we obtain the chiral kinetic equation for antiparticles,
\begin{eqnarray}
 \left(1 - s\hbar B\cdot \Omega\right)  n\cdot \partial^x  \bar f_{\bar p}^{\textrm t} & &\nonumber\\
+ \left[v^\mu - s\hbar ( \hat{\bar p} \cdot \Omega ) B^\mu
- {s\hbar}\epsilon^{\mu\nu \rho \sigma} n_\rho E_\sigma \Omega_\nu\right] \bar \partial^x_\mu \bar f_{\bar p}^{\textrm t} & &\nonumber\\
-\left( \tilde E^{\mu} + \epsilon^{\mu\nu\alpha\beta} v_{\nu} n^\alpha B^\beta
 - {s\hbar} E\cdot B    \Omega_\mu \right)\bar \partial^\mu_p   \bar f_{\bar p}^{\textrm t} & &  \nonumber\\
 + {s\hbar}E\cdot B \left( \bar\partial^\mu_p \Omega_\mu\right) \bar f_{\bar p}^{\textrm t} &=&0, \hspace{1cm}
\end{eqnarray}
where
\begin{eqnarray}
 \bar f_{\bar p}^{\textrm t}&=&\left.f\left(p_n=-|\bar p|(1-\hbar s B\cdot \Omega)\right)\right|_{\bar p=-\bar p,\, s=-s,}\\
v^\mu &=& \left[ \left(1 - \frac{s\hbar B\cdot \bar p}{|\bar p|^3}\right)\hat{\bar p}^\mu
- \frac{s\hbar B^\mu}{2|\bar p|^2}\right].
\end{eqnarray}
Here, $\bar f_{\bar p}^t =\bar f_{\bar p} + \bar f_{\textrm v}$ denotes the total contribution
of the normal $\bar f_{\bar p}$ and the vacuum distribution $\bar f_{\textrm v}$.
Using the free quantum Dirac field theory with $\bar f_\textrm{v}=-1$,
we obtain the kinetic equation of the normal distribution $\bar f_{\bar p}$ as,
\begin{eqnarray}
\label{cke-antiparticle}
 \left(1 - s\hbar B\cdot \Omega\right)  n\cdot \partial^x  \bar f_{\bar p} & &\nonumber\\
+ \left[v^\mu - s\hbar ( \hat{\bar p} \cdot \Omega ) B^\mu
- {s\hbar}\epsilon^{\mu\nu \rho \sigma} n_\rho E_\sigma \Omega_\nu\right] \bar \partial^x_\mu \bar f_{\bar p} & &\nonumber\\
-\left( \tilde E^{\mu} + \epsilon^{\mu\nu\alpha\beta} v_{\nu} n^\alpha B^\beta
 - {s\hbar} E\cdot B    \Omega_\mu \right)\bar \partial^\mu_p   \bar f_{\bar p} & &  \nonumber\\
 + {s\hbar}E\cdot B \left( \bar\partial^\mu_p \Omega_\mu\right)\left( \bar f_{\bar p} -1\right) &=&0. \hspace{1cm}
\end{eqnarray}
It should be noted that it is sufficient to use the result $\bar f_\textrm{v}=-1$ from the free quantum field here
in order to describe the chiral anomaly because the chiral anomaly term with $E\cdot B$ in Eq.~(\ref{cke-antiparticle})
always arises at the first order of $\hbar$.
The last term in Eq.~(\ref{cke-antiparticle}) is different from that in Eq.~(\ref{cke-particle}) for particle due to the Dirac sea contribution.
There exists an inhomogeneous term
$-{s\hbar}E\cdot B \left( \bar\partial^\mu_p \Omega_\mu\right)$ originated  from the Dirac sea.
As discussed in the last section, this term
 will eventually lead to  the chiral anomaly.
 In the scenario of the Dirac sea, the difference between the particle and antiparticle is very natural.

\section{Summary}

Starting from the quantum transport theory based on the quantum field theory,
we find out that the Wigner equation holds only for Wigner functions that are not normal ordered.
It turns out that the Dirac sea or the vacuum contribution originated from the anti-commutation relation
between antiparticle field operators in un-normal-ordered Wigner function plays a central role
in generating the right chiral anomaly for  both  massive fermions and  massless fermions.
Correspondingly, the  chiral kinetic equation for antiparticles should include the contribution from the Dirac sea contribution.

\section{Acknowledgments}

We thank Pengfei Zhuang and Jian Zhou for insightful discussions. 
This work was supported in part by the National Natural Science Foundation of China under
Nos. 11890713 and 11675092, and the Natural Science Foundation of Shandong Province under No. JQ201601.

\appendix

\section{Integral calculation}

In this appendix, we give the details on how to calculate the normal contribution from $f^\pm$ and $\bar f^\pm$  in  Eq.~(\ref{CA_C}).

Substituting the normal distribution function given by  Eq.~(\ref{Fpm}) into Eq.~(\ref{CA_C}) and using the identities,
\begin{eqnarray}
& &\delta'\left(p^2-m^2\right)\nonumber\\
&&=\frac{1}{4p_0 E_p }\left[\frac{}{}\delta'\left(p_0-E_p\right)+\delta'\left(p_0+E_p\right)\right]\nonumber\\
&&=\frac{1}{4E_p^3}\left[\frac{}{}\delta\left(p_0-E_p\right)+\delta\left(p_0+E_p\right)\right]\nonumber\\
&&~+\frac{1}{4E_p^2}\left[\frac{}{}\delta'\left(p_0-E_p\right)-\delta'\left(p_0+E_p\right)\right], \hspace{1cm}
\end{eqnarray}
we decompose Eq.~(\ref{CA_C}) into  the following three parts,
\begin{eqnarray}
\label{P-2-n-int0}
C_{\textrm n}&=&C_{\textrm n}^{(1)}+C_{\textrm n}^{(2)}+C_{\textrm n}^{(3)},
\end{eqnarray}
\begin{eqnarray}
\label{Cn1}
C_{\textrm n}^{(1)} &=&- \int \frac{ d^4 p}{8\pi} {\mathbf \partial}_{\bf p}\cdot \left[\frac{\bf p}{E_p^3}  (f^+ + f^-) \delta\left(p_0 - E_p\right)\right]\nonumber\\
&-& \int \frac{ d^4 p}{8\pi} {\mathbf \partial}_{\bf p}\cdot \left[\frac{\bf p}{E_p^3}  (\bar f^+ + \bar f^-) \delta\left(p_0 + E_p\right)\right], \\
C_{\textrm n}^{(2)} &=&- \int \frac{ d^4 p}{8\pi} {\mathbf \partial}_{\bf p}\cdot \left[\frac{\bf p}{E_p^2}  (f^+ + f^-) \delta^{\prime}\left(p_0 - E_p\right)\right]\nonumber\\
& +& \int \frac{ d^4 p}{8\pi} {\mathbf \partial}_{\bf p}\cdot \left[\frac{\bf p}{E_p^2}  (\bar f^+ + \bar f^-) \delta^{\prime}\left(p_0 + E_p\right)\right], \\
%
C_{\textrm n}^{(3)} &=&- \int \frac{d^4 p}{4\pi}  \partial_{p_0} \left[ \frac{1}{E_p}   (f^+ + f^-)\delta^{\prime}\left(p_0-E_p\right)\right]\nonumber\\
&- & \int \frac{d^4 p}{4\pi}  \partial_{p_0} \left[ \frac{1}{E_p}   (\bar f^+ + \bar f^-)\delta^{\prime}\left(p_0+E_p\right)\right].
\end{eqnarray}
For $C_{\textrm n}^{(1)}$ and $C_{\textrm n}^{(2)}$, we  integrate $p_0$ over the delta function or derivative of delta function  and obtain,
\begin{eqnarray}
C_{\textrm n}^{(1)} &=&- \int \frac{ d^3 {\bf p}}{4\pi} {\mathbf \partial}_{\bf p}\cdot \left[\frac{\bf p}{E_p^3}( f_{\bf p}+\bar f_{\bf p}) \right],\\
C_{\textrm n}^{(2)} &=&- \int \frac{ d^3 {\bf p}}{4\pi} {\mathbf \partial}_{\bf p}\cdot \left[\frac{\bf p}{E_p^2} ( f'_{\bf p} +\bar f'_{\bf p}) \right],
\end{eqnarray}
where
\begin{eqnarray}
\label{def:f}
f_{\bf p} &=&\frac{1}{2} \left. (f^++f^-)\right|_{p_0=E_p},\nonumber\\
\bar f_{\bf p} &=& \frac{1}{2} \left. (\bar f^+ + \bar f^-) \right|_{p_0=-E_p},\nonumber\\
f'_{\bf p} &=& -\frac{1}{2} \left. \partial_{ p_0}(f^++f^-)\right|_{p_0=E_p},\nonumber\\
\bar f'_{\bf p}&=& \frac{1}{2} \left. \partial_{ p_0}(\bar f^+ + \bar f^-) \right|_{p_0=-E_p}.
\end{eqnarray}
It is obvious that both $C_{\textrm n}^{(1)}$ and $C_{\textrm n}^{(2)}$ vanish for the normal distribution function
that approaches  zero rapidly at infinity in momentum space.

For $C_{\textrm n}^{(3)}$, it is more convenient to integrate $|{\bf p}|$ over the derivative of delta function,
\begin{eqnarray}
C_{\textrm n}^{(3)}
&=&- \int \frac{d\Omega d p_0}{8\pi} \partial_{p_0}^2 \left[ \sqrt{p_0^2-m^2}f_{p_0} \theta(p_0-m)\right] \nonumber\\
&- & \int \frac{d\Omega d p_0}{8\pi} \partial_{p_0}^2 \left[ \sqrt{p_0^2-m^2} \bar f_{p_0}\theta(-p_0-m) \right] \nonumber\\
&+& \int \frac{d\Omega d p_0}{8\pi} \partial_{p_0}\left[\sqrt{p_0^2-m^2} f'_{p_0} \theta(p_0-m) \right]\nonumber\\
&+ & \int \frac{d\Omega d p_0}{8\pi} \partial_{p_0}\left[\sqrt{p_0^2-m^2}\bar f'_{p_0} \theta(-p_0-m) \right], ~~~
\end{eqnarray}
where
\begin{eqnarray}
f_{p_0} &=& \frac{1}{2}\left. (f^++f^-)\right|_{|{\bf p}|=\sqrt{p_0^2-m^2}},\\
\bar f_{p_0} &=& \frac{1}{2} \left. (\bar f^+ + \bar f^-) \right|_{|{\bf p}|=\sqrt{p_0^2-m^2}},\\
f'_{p_0} &=& \frac{1}{2}\left. \partial_{p_0}(f^++f^-)\right|_{|{\bf p}|=\sqrt{p_0^2-m^2}},\\
\bar f'_{p_0}&=& \frac{1}{2} \left.  \partial_{p_0} (\bar f^+ + \bar f^-)\right|_{|{\bf p}|=\sqrt{p_0^2-m^2}},
\end{eqnarray}
and $\theta(\pm p_0-m )$ is step function.
It is obvious that the last two terms vanish because the function in square brackets
vanishes at  the boundary points $p_0= \pm m,\pm \infty$.
For the first two terms, when one of the derivative acts on $f_{p_0}$ or step function, the integral also vanishes.
However when the derivative acts on $\sqrt{p_0^2-m^2}$, the reciprocal of $\sqrt{p_0^2-m^2}$ will
arise and leads to possible  divergence at boundary point $p_0=\pm m$,
\begin{eqnarray}
C_{\textrm n}^{(3)}
&=&- \int \frac{d\Omega d p_0}{4\pi} \partial_{p_0} \left[ \frac{p_0 f_{p_0}}{\sqrt{p_0^2-m^2}} \theta(p_0-m)\right] \nonumber\\
&-& \int \frac{d\Omega d p_0}{4\pi} \partial_{p_0} \left[\frac{p_0  \bar f_{p_0}}{ \sqrt{p_0^2-m^2}}\theta(-p_0-m)\right]. ~~~~~~
\end{eqnarray}
We expand each term above according to whether  the derivative acts on step function or not,
\begin{eqnarray}
C_{\textrm n}^{(3)}
&=&- \int \frac{d\Omega }{4\pi} \int_m^{+\infty} d p_0  \partial_{p_0} \left( \frac{p_0 f_{p_0} }{\sqrt{p_0^2-m^2}}\right) \nonumber\\
&- & \int \frac{d\Omega d p_0}{4\pi} \frac{p_0 f_{p_0}}{\sqrt{p_0^2-m^2}} \delta(p_0-m)\nonumber\\
& -& \int \frac{d\Omega }{4\pi} \int_{-\infty}^{-m} d p_0\partial_{p_0} \left(\frac{p_0 \bar f_{p_0}}{ \sqrt{p_0^2-m^2}}  \right)\nonumber\\
&+ & \int \frac{d\Omega d p_0}{4\pi} \frac{p_0 \bar f_{p_0}}{\sqrt{p_0^2-m^2}} \delta(p_0+m).
\end{eqnarray}
In order to regularize the divergence at $p_0=m$ and  $p_0=-m$, we set $p_0=m+\epsilon$ and $p_0=-m-\epsilon$, respectively.
It follows that

\begin{eqnarray}
\label{Cn1-final}
&&C_{\textrm n}^{(3)}=
\frac{m+\epsilon }{\sqrt{(2m+\epsilon) \epsilon}}\int \frac{d\Omega }{4\pi}
\left( f_{p_0=m+\epsilon} + \bar f_{p_0=-m-\epsilon} \right)\nonumber\\
&& -\frac{m+\epsilon }{\sqrt{(2m+\epsilon) \epsilon}}\int \frac{d\Omega }{4\pi}
\left(f_{p_0=m+\epsilon} + \bar f_{p_0=-m-\epsilon} \right).~~~
\end{eqnarray}
We see that each term is divergent at the limit $\epsilon\rightarrow 0$ when $m\neq 0$,
but every two terms cancel each other and give null result.
It is interesting that for the chiral limit $m=0$ there is no divergence  and we obtain,
\begin{eqnarray}
\label{Cn1-final}
C_{\textrm n}^{(3)}&=&
\int \frac{d\Omega }{4\pi}\left( f_{p_0=0}+\bar f_{p_0=0}\right) \nonumber\\
& &-\int \frac{d\Omega }{4\pi}\left( f_{p_0=0}+\bar f_{p_0=0}\right)
\end{eqnarray}
As has been mentioned in Sec.~\ref{sec:chiral}, the first term happens to be the right coefficient of the chiral anomaly
if  we choose the Fermi-Dirac distribution function given by Eq.~(\ref{Fermi-Dirac}).

\end{document}